\documentclass[fleqn,usenatbib]{mnras}

\usepackage{newtxtext,newtxmath}

\usepackage[T1]{fontenc}

\DeclareRobustCommand{\VAN}[3]{#2}
\let\VANthebibliography\thebibliography
\def\thebibliography{\DeclareRobustCommand{\VAN}[3]{##3}\VANthebibliography}
\def\BF{\ensuremath{\mathcal{B}}}

\usepackage{graphicx}	
\usepackage{amsmath}	
\graphicspath{{./}{plots/}}
\usepackage{tabularx}

\title{Revisiting the Properties of GW190814 and Its Formation History}
\title[Understanding the origin of GW190814]{Revisiting the Properties of GW190814 and Its Formation History}
\author[F.\, Lyu et al.]{F.\, Lyu,$^{1}$
L.\, Yuan,$^{2}$
D. H.\, Wu,$^{2}$
W. H.\, Guo,$^{3}$
Y. Z.\, Wang,$^{3}$
S. X.\, Yi,$^{4}$
Q. W.\, Tang,$^{5}$\thanks{E-mail: qwtang@ncu.edu.cn}
R.-C.\, Hu,$^{6}$
J.-P.\, Zhu,$^{7,8}$
\newauthor
X. W.\, Shu,$^{2}$
Y.\, Qin$^{2,6}$\thanks{E-mail: yingqin2013@hotmail.com}
and E. W.\, Liang$^{6}$\thanks{E-mail: lew@gxu.edu.cn} \\
$^{1}$Astronomical Research Center, Shanghai Science \& Technology Museum, Shanghai 201306, China\\
$^{2}$Department of Physics, Anhui Normal University, Wuhu, Anhui 241002, China\\
$^{3}$Key Laboratory of Dark Matter and Space Astronomy, Purple Mountain Observatory, Chinese Academy of Sciences, Nanjing 210033, China\\
$^{4}$School of Physics and Physical Engineering, Qufu Normal University, Qufu, Shandong 273165, China\\
$^{5}$Department of Physics, School of Physics and Materials Science, Nanchang University, Nanchang 330031, China\\
$^{6}$Guangxi Key Laboratory for Relativistic Astrophysics, School of Physical Science and Technology, Guangxi University, Nanning 530004, China\\
$^{7}$School of Physics and Astronomy, Monash University, Clayton Victoria 3800, Australia\\
$^{8}$OzGrav: The ARC Centre of Excellence for Gravitational Wave Discovery, Clayton Victoria 3800, Australia\\
}

\date{Accepted 2023 August 18. Received 2023 August 15; in original form  2023 July 4}

\begin{document}
\label{firstpage}
\pagerange{\pageref{firstpage}--\pageref{lastpage}}
\maketitle

\begin{abstract}
GW190814 was reported during LIGO’s and Virgo’s third observing run with the most asymmetric component masses (a $\sim 23$ $M_{\odot}$ black hole and a $\sim2.6$ $M_{\odot}$ compact object). Under the assumption that this event is a binary black hole (BBH) merger formed through the isolated binary evolution channel, we reanalyze the publicly released data of GW190814 with the modified astrophysical priors on the effective spin $\chi_{\rm eff}$, and further explore its formation history using detailed binary modeling. We show that GW190814 is likely to have been formed through the classical common envelope channel. Our findings show that the properties inferred using the modified astrophysical priors are consistent with those inferred by the uniform priors. With the newly-inferred properties of GW190814, we perform detailed binary evolution of the immediate progenitor of the BBH (namely a close binary system composed of a BH and a helium star) in a large parameter space, taking into account mass-loss, internal differential rotation, supernova kicks, and tidal interactions between the helium star and the BH companion. Our findings show that GW190814-like events could be formed in limited initial conditions just after the common envelope phase: a $\sim 23$ $M_{\odot}$ BH and a helium star of $M_{\rm ZamsHe}$ $\sim$ 8.5 $M_{\odot}$ at solar metallicity ($\sim$ 7.5 $M_{\odot}$ at 10\% solar metallicity) with an initial orbital period at around 1.0 day. Additionally, the inferred low spin of the secondary indicates that the required metallicity for reproducing GW190814-like events should not be too low (e.g., Z $\gtrsim$ 0.1 $Z_{\odot}$).
\end{abstract}

\begin{keywords}
 Stars:Black holes -- Stars:binaries -- Stars: Wolf-Rayet -- Gravitational waves
\end{keywords}



\section{Introduction}
The gravitational wave (GW) event GW190814 was reported by the LIGO
Scientific and Virgo Collaborations (LVC) with a signal-to-noise ratio of 25 in the three-detector network during the third observing run on 2019 August 14 \citep{Abbott2020a}. This event stands out with the most unequal mass ratio of $q = m_2/m_1 = 0.112^{+0.008}_{-0.009}$ (all measurements are reported at 90\% confidence level unless otherwise specified). The primary component is conclusively a black hole (BH) with mass $m_1 = 23.2^{+1.1}_{-1.0} M_{\odot}$, while the nature of the secondary ($m_2 = 2.59^{+0.08}_{-0.09}M_{\odot}$) is still unclear with its mass in the hypothesized lower mass gap of 2.5 - 5 M$_{\odot}$ \citep{Bailyn1998,Ozel2010,Farr2011,Ozel2012,Fryer2012,Zevin2020a,Safarzadeh2020,Zhu2021,Shao2022,Olejak2022,Siegel2022}. The dimensionless spin of the secondary component ($\chi_2$) remains unconstrained, however, the primary BH spin is tightly constrained down to $\chi_1$ $\leqslant$ 0.07. This is the strongest constraint on the primary spin for any event so far \citep{Abbott2019,Abbott2020b,Abbott2021b}, which is consistent with the predictions of isolated binary evolution scenario \citep{Qin2018,Fuller2019,Bavera2020,Belczynski2020}. 

The origin of GW190814 is still under debate, although different groups have independently investigated the formation scenario of GW190814. With a modified supernova (SN) engine model, \cite{Zevin2020b} found it still challenging for population synthesis modeling \citep{Breivik2020} to match the formation rate of GW190814-like events. Subsequently, \cite{Safarzadeh2020} proposed an alternative scenario, in which the first-born NS could accrete a large fraction of the ejecta mass due to the presence of a massive BH companion from the accretion disk and transitioned into becoming a mass gap object. \cite{Kinugawa2021} performed population synthesis simulations of Population III binary stars to find the existence of GW190814-like events. More recently, \cite{Antoniadis2022} suggested that a sizable fraction of stars with pre-collapse carbon-oxygen (CO) core masses in the range of 10 - 30 $M_{\odot}$ could form GW190814-like events.

In addition, other formation scenarios of GW190814 include hierarchical mergers in multiple systems \citep{Lu2021,Liu2021} or in active galactic nuclei \citep{Yang2020,Tagawa2021}, and dynamical encounters in young star clusters \citep{Rastello2020} or in certain globular cluster environments \citep{Kritos2021}. Alternatively, GW190814 event could be primordial BHs \citep{Clesse2020,Franciolini2022,Jedamzik2021,Chen2022,Escriva2023}. \cite{Vattis2020}, however, pointed out that a pair of primordial- and stellar-origin BH binaries for GW190814 is unlikely since it is limited by such a binary to form and merge within a Hubble time.  

It has been pointed out in recent studies \citep{Roulet2021,Galaudage2021,Payne2022,Qin2022GW190403} that the inferred properties of BBHs are heavily dependent on the choice of priors. Several groups independently investigated the origin of some specific GW events with different priors. Within the context of isolated binary evolution and assuming the first-born more massive BH with a negligible spin, \cite{Mandel2020} reweighed the $\chi_{\rm eff}$ posterior samples from the \texttt{EOB} analysis to conclude that the less massive BH was spinning faster than the one inferred by the LVC using the uninformative priors. Subsequently, with no such restrictions on spins, \cite{Zevin2020b} re-analyzed the data from both the \texttt{Phenom} and \texttt{EOB} using a suite of priors from various astrophysical predictions to reach constraints on the secondary spin similar to \cite{Mandel2020}, but with a slightly different mass ratio. \cite{Mandel2021} suggested that 
a prior of nonspinning BH for GW200115 is more consistent with current astrophysical understanding. As for GW190521, a massive BBH merger reported in \cite{GW190521}, \cite{Fishbach2020} adopted a population-informed prior to re-analyze its data, concluding that this event is more likely the first event with two component masses straddling the pair-instability gap. With the population-informed priors, \cite{Qin2022GW190403} derived the re-weighted posterior distributions for a subsample of 52 BBH events (see their Figure 4). Recently, reanalyzing the second LIGO-Virgo catalog of BBHs with a range of isolated and dynamical formation channels, \citet{Zevin2021} found that multiple formation channels are at play. In the context of an isolated binary evolution channel, \cite{Qin2022CHE} performed Bayesian inference for all BBH events with various astrophysical priors \citep{Zevin2021} representing different formation channels (common-envelope evolution channel, CEE; stable mass transfer, SMT; chemically homogeneous evolution, CHE). Their findings show that GW190517 is most likely to have been formed through the CHE channel. 

In this work, assuming BBHs formed through an isolated binary evolution scenario, we first diagnose the formation channel of GW190814 and then perform Bayesian inference using the astrophysical priors on the properties of GW190814. With the newly-inferred parameters (masses and spins of the two BH components), we further adopt the detailed binary evolution to investigate the formation history of GW190814-like events. In Section ~\ref{sect2}, we infer the properties of GW190814 with the astrophysical priors. We then perform detailed binary evolution to systematically investigate the formation history of GW190814-like events in Section ~\ref{sect3}. Our results for detailed binary evolution are presented in Section ~\ref{sect4}. Finally, the main conclusions and some discussion are summarized in Section ~\ref{sect5}.

\section{Properties of GW190814 inferred with the astrophysical priors}\label{sect2}

The Bayes factor indicates whether any one prior assumption is favored or disfavored by the data compared to another \citep{Zevin2020a}. Following the approaches in \cite{Qin2022CHE}, we adopt the different astrophysical priors \citep{Zevin2021} representing BBH formation channels of isolated binary evolution (CEE, SMT, and CHE) in the Bayesian inference to calculate the corresponding Bayes evidence and study the influences of the priors on the inferred properties of BBHs. The resulting Bayes factors between the priors are listed in Table 1. We find that the data prefer the CEE prior over the SMT prior with modest support ($\BF \sim 7$), and the CHE prior is strongly disfavored. 

We note that all three astrophysical models are expected to produce BBHs with comparable masses \citep[see Figure 1 in][]{Qin2022CHE}, and the mass ratio distributions as predicted by these models have brought strong information that affects the posterior distributions for the parameters of GW190814 in the inferences: we have inferred $q > 0.2$ for all three priors. For each of the inferences adopting the astrophysical priors, we find the parameter set having the largest likelihood and compare it with the one inferred with the LIGO-Virgo-KAGRA (LVK) default priors, i.e., uniform distributions for the detector-frame component masses, which infers $q \sim 0.1$. The results are demonstrated in the second row of Table 1, and it shows that the maximum likelihood value using the LVK default prior ($\mathcal{L}_0$) is greater than that of the astrophysical priors ($\mathcal{L}$). 

The results above indicate that while the astrophysical priors can provide clues for the judgment of formation channel, for a minority of events in the population (like GW190814), these priors may dominate the inference on the mass distributions and drive the results away from the parameter space that can best-fit the observed strain data. In order to obtain posterior distributions that are mainly constrained by the data, in the following we neglect the predictions on the component masses from the astrophysical models and use only the predictions on the spins as priors to derive the parameters for this event.

\begin{table}
\centering
\caption{The Bayes factors and the maximum likelihoods for the astrophysical priors (see Figure 1 in \citet{Qin2022CHE} for detailed illustrations of different astrophysical priors).}
\label{tab:your_label}
\begin{tabular}{cccc} 
\hline
 & CEE & SMT & CHE \\
\hline
$\ln \BF$ & 0 & -1.96  & -86.19 \\
$\ln(\mathcal{L}/\mathcal{L}_0)$ & -164.94 & -173.38 & -265.94 \\
\hline
\end{tabular}
\end{table}

For the LVK default priors on BH spins, \citet{Abbott2020b} assumed that they are uniform in magnitudes $\chi_i$ ($i=1,2$) and isotropic in directions. In the classical CEE scenario, it is widely expected that the two BHs are aligned with each other and to the total orbital
angular momentum. The effective inspiral spin ($\chi_{\rm eff}$), which can be directly constrained by the GW signal, is defined as follows:

\begin{equation}\label{eq1}
    \chi_{\rm eff}=\frac{m_1\chi_{1z} + m_2\chi_{2z}}{m_1+m_2},
\end{equation}
where $m_1$ and $m_2$ are the masses of the primary and secondary BH, and $\chi_{1z}$ and $\chi_{2z}$ are the corresponding BH spin magnitudes aligned to the direction of the orbital angular momentum.
 
\begin{figure}
    \centering    
    \includegraphics[width=\columnwidth]{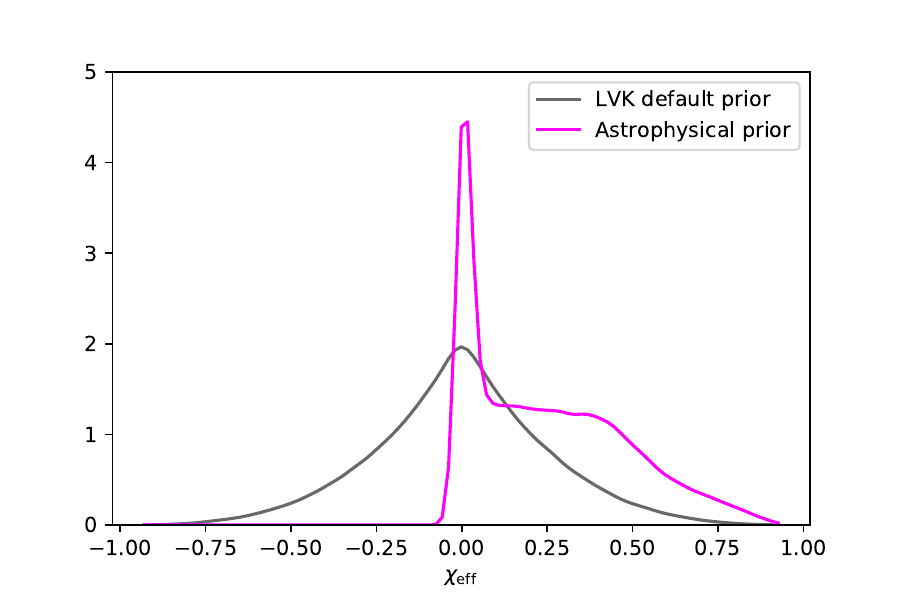}
    \caption{The prior of $\chi_{\rm eff}$ from \citet{Abbott2020a} (shown in gray) and our astrophysical prior on $\chi_{\rm eff}$ (shown in magenta).} 
    \label{fig:chi_eff_prior}
\end{figure}

In this section, we describe in detail which priors for both BH spins and masses are chosen for further inferring the properties of GW190814. As demonstrated above, GW190814 is most likely to be formed through the CEE channel. Assuming the highly efficient transport of the angular momentum \citep[e.g.,][]{Fuller2019} between the stellar core and its envelope, the primary star (initially a more massive star on zero-age main sequence) is expected to form a BH (the first-born BH) with negligible spin \citep{Qin2018}. We also note that a slightly higher value of BH spin ($\sim 0.1$) could be reached in \cite{Belczynski2020}. In order to obtain the distribution of $\chi_{\rm eff}$, we adopt: i) the LVK default uniform priors on the masses of two components; ii) astrophysically-predicted non-spinning first-born BH and $\chi_{\rm 2z}$ predicted in \cite{Zevin2021} \footnote{We adopt $\chi_{\rm 1z} = 0$ for the non-spinning first-born BH and the $\chi_{\rm 2z}$ distribution in Figure 1 of \citet{Qin2022CHE}.}. We present the two distributions of $\chi_{\rm eff}$ in Figure \ref{fig:chi_eff_prior}. We do not adopt BH mass distributions of \cite{Zevin2021} as our mass priors, as such a low mass ratio of GW190814 was predicted with a very low probability \citep[see the solid blue line on the top right panel in their Figure 1 in][]{Qin2022CHE}. Additionally, we also do not consider the population-informed priors on BH masses. This is because GW190814 was found to be an outlier \citep{Abbott2023pop} in the detailed analysis of the mass distribution of the BBH population. Therefore, we combine a hybrid prior, the astrophysical predictions of $\chi_{\rm eff}$ and the uniform BH mass distribution (LVK default priors), to further infer the properties of GW190814. 

\begin{figure*}
    \centering
    \includegraphics[width=2\columnwidth]{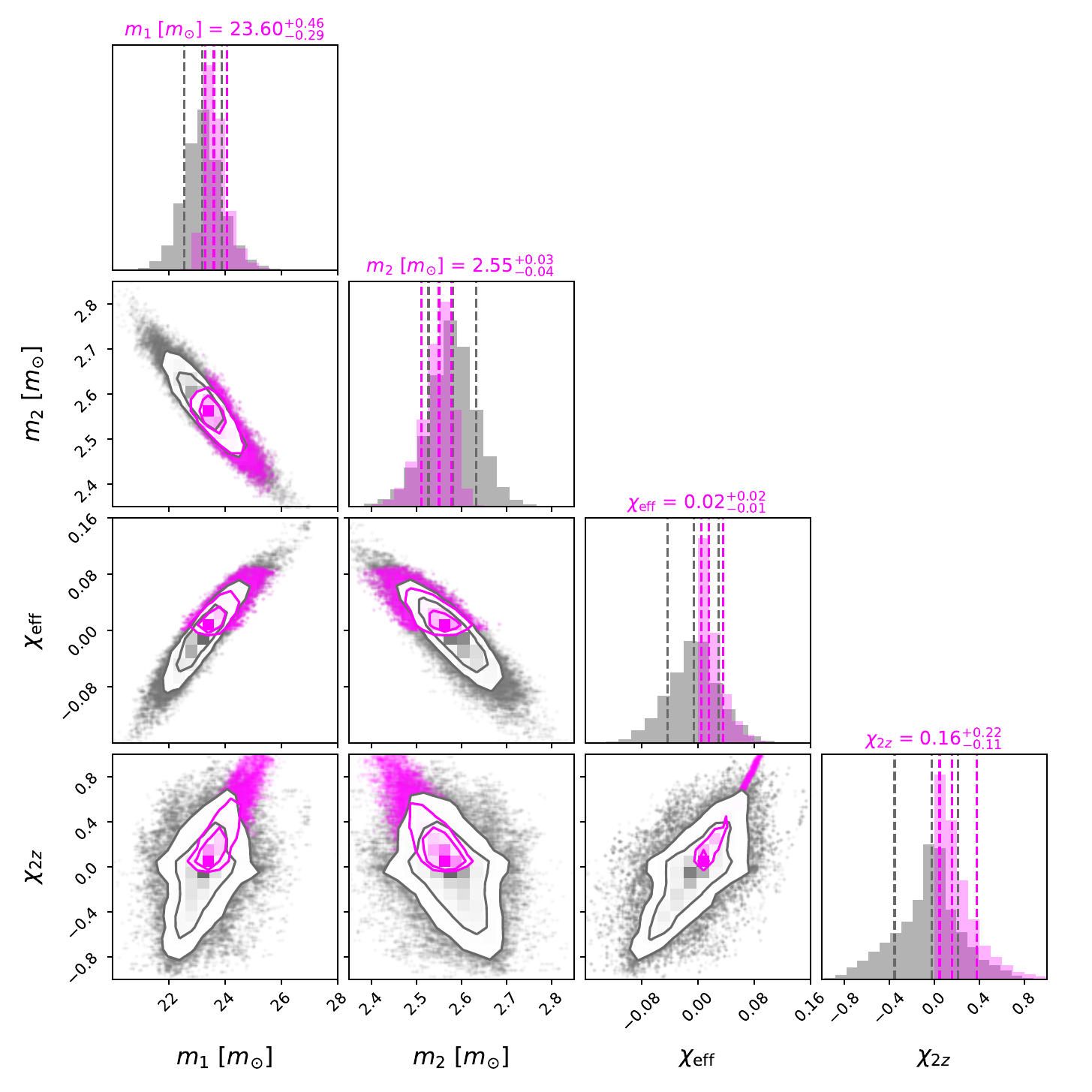}
    \caption{Corner plot of the posterior distributions of $m_1$, $m_2$, $\chi_{\rm eff}$ and $\chi_{2z}$ from LIGO (gray) and this work (magenta). Contours show $68\%$ and $99\%$ credible regions. The corresponding marginal posterior medians and one-sigma errors are also shown.} 
    \label{fig:posterior_compare}
\end{figure*}

Following the method discussed in \citet{Mandel2020}, we re-weight the posterior samples from \citet{Abbott2020b} (The IMRPhenomPv3HM waveform) using the astrophysical prior on $\chi_{\rm eff}$. The posteriors on $m_1$, $m_2$, $\chi_{\rm eff}$ and $\chi_{2z}$ under this modified prior are shown in Figure \ref{fig:posterior_compare}. The updated constraints, based on our new astrophysical $\chi_{\rm eff}$ prior, are consistent with those inferred by the LVK default priors. It is worth noting that the newly inferred $\chi_{\rm eff}$ and $\chi_{2z}$ are shifted to slightly larger values. Additionally, we obtain a higher median mass for the primary and lower one for the secondary (see Figure \ref{fig:posterior_compare}) when compared with those from \citet{Abbott2020b}, leading to a more extreme mass ratio ($q = m_2 /m_1$). This is partly due to the $\chi_{\rm eff} - q$ anticorrelation, which has also been demonstrated in earlier studies \cite[e.g.,][]{Mandel2021}. We assume that a compact object with its mass not smaller than 2.5 $M_{\odot}$ is considered to be a BH throughout our work. Therefore, GW190814 is assumed to be a BBH merger event. 

\section{Binary evolutionary history of GW190814}\label{sect3}
\subsection{Methods}
Assuming GW190814 formed through the CEE scenario, the immediate progenitor of the BBH is a close binary system composed of a BH and a helium star. We here perform detailed binary modeling using release 15140 of the \texttt{MESA} stellar evolution code \citep{Paxton2011,Paxton2013,Paxton2015,Paxton2018,Paxton2019,paxton2023} to reveal its formation history. We follow the same method in previous studies \citep{Qin2018,Qin2023,Bavera2020,Hu2022,Hu2023,Fragos2023} to first create pure helium stars at the zero-age helium main sequence (ZamsHe) in a mass range of 2.5 - 10 $M_{\odot}$ and then relax ZamsHe to reach the thermal equilibrium where the luminosity from helium burning just exceeds 99\% of the total luminosity. We adopt $Z_{\odot} = $ 0.0142 as the solar metallicity \citep{Asplund2009}. 

We model convection using the mixing-length theory \citep{MLT1958} with a $\alpha_{\rm mlt} =$ 1.93. We adopt the Ledoux criterion to treat the boundaries of the convective zone and consider the step overshooting as an extension given by $\alpha_p=0.1H_p$, where $H_p$ is the pressure scale height at the Ledoux boundary limit. Semiconvection \citep{Langer1983} with an efficiency parameter $\alpha_{\sc}=1.0$ is adopted in our modeling. The network of \texttt{approx12.net} is chosen for nucleosynthesis. We treat rotational mixing and angular momentum transport as diffusive processes \citep{Heger2000}, including the effects of the Goldreich–Schubert–Fricke instability, Eddington–Sweet circulations, as well as secular and dynamical shear mixing. For an efficient angular momentum transport within stars, we adopt the Spruit-Tayler dynamo \cite[e.g.,][]{Spruit1999,Spruit2002}. We adopt diffusive element mixing from these processes with an efficiency parameter of $f_c=1/30$ \citep{Chaboyer1992, Heger2000}. 

Stellar winds of helium stars are modeled following the same method as in \cite{Hu2022}. We evolve helium stars in binaries until the depletion of their central carbon, from which the baryonic remnant mass is then calculated following the “delayed” supernova prescription \citep{Fryer2012}. We also take into account the neutrino loss as in \cite{Zevin2020b}. The maximum mass of an NS is 2.5 $M_{\odot}$. We calculate the timescale for orbital synchronization
following \cite{Hurley2002} for massive
stars with radiative envelopes and adopted the updated fitting formula for the tidal coefficient $E_2$ provided in \cite{Qin2018}. We assume the Eddington-limited accretion onto BHs using the standard formulae \citep[e.g.,][]{Frank2002,Antoniadis2022,Qin2022RAA,Fragos2023}.

\section{Results}\label{sect4}
\begin{figure*}
     \centering
     \includegraphics[width=1.0\textwidth]{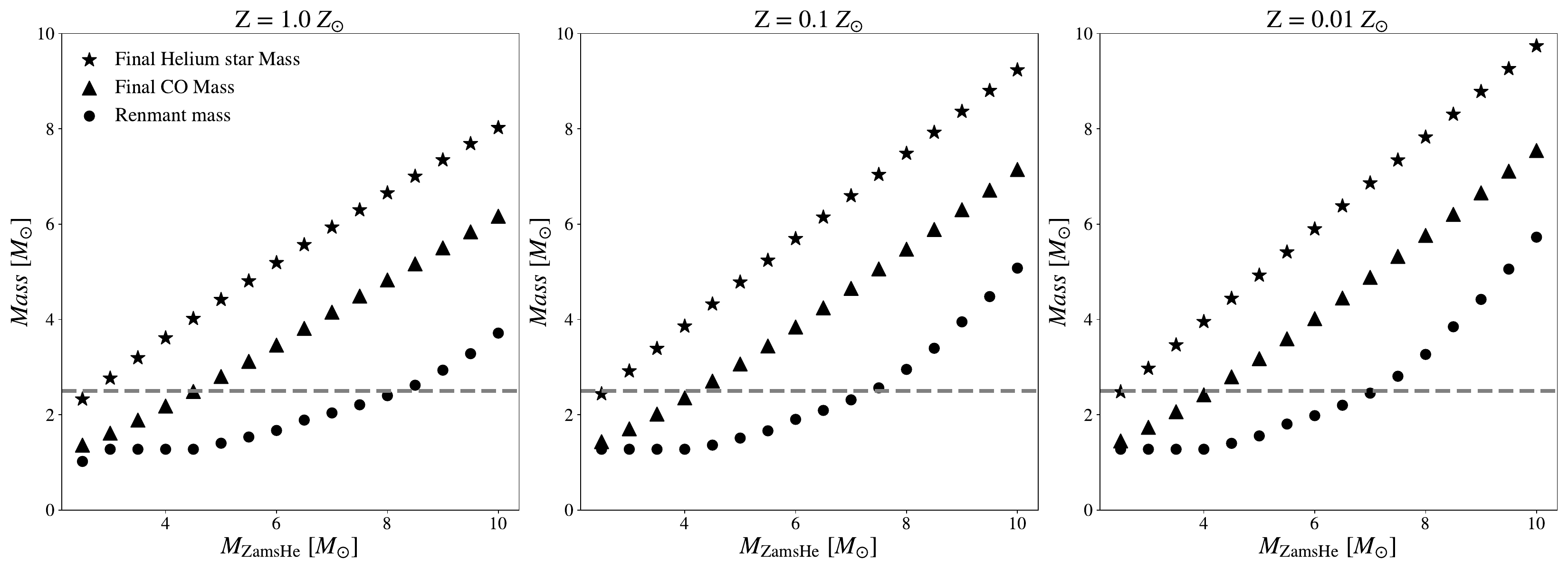}
     \caption{Helium star's final mass (star symbol) and its CO core (triangle) at central carbon depletion and remnant mass (dot) as a function of the helium star initial mass in the range of 2.5 - 10 $M_\odot$. Left panel: 1.0 $Z_\odot$, middle panel: 0.1 $Z_\odot$, right panel: 1.0 $Z_\odot$. The horizontal line represents the remnant mass of 2.5 $M_{\odot}$.}
     \label{fig:mass}
\end{figure*}

We present in Figure \ref{fig:mass} the evolutionary results of single nonrotating helium stars with ZamsHe mass from 2.5 to 10 $M_{\odot}$ at three different metallicities. First, the final masses of helium stars and their carbon-oxygen (CO) core increase linearly with the ZamsHe mass for different metallicities (1.0 $Z_{\odot}$, 0.1 $Z_{\odot}$ and 0.01 $Z_{\odot}$), while the corresponding remnant mass increases slowly when the initial helium star has a mass below 8 $M_{\odot}$. Second, we can see that metallicity plays a significant role in the evolution of helium stars. This is because helium stars tend to lose more mass due to metallicity-dependent winds \citep{Vink2005,Eldridge2006,Sander2020}. Additionally, we also note that the threshold for a helium star at 1.0 $Z_{\odot}$ evolving to form a BH is around 8 $M_{\odot}$ (7.5 $M_{\odot}$ at 0.1 $Z_{\odot}$ and 7 $M_{\odot}$ at 0.01 $Z_{\odot}$).

\begin{figure*}
     \centering
     \includegraphics[width=1.0\textwidth]{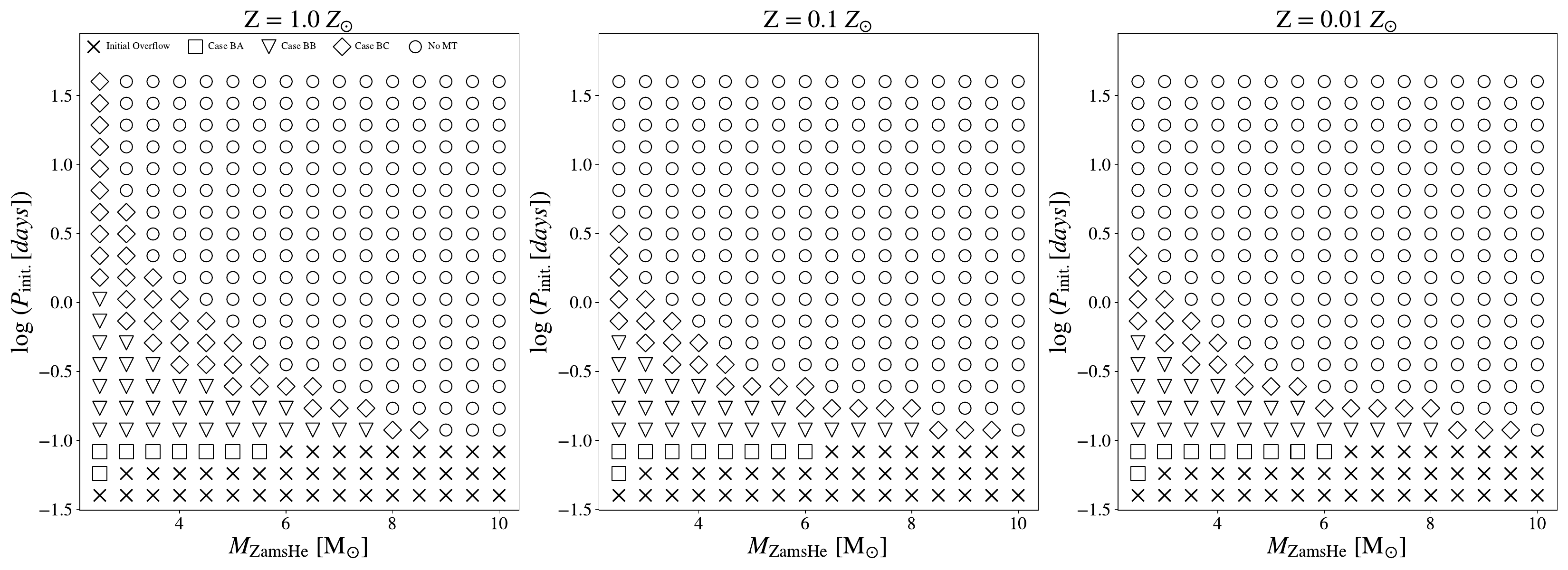}
     \caption{Various interactions of a helium star (Left panel: 1.0 $Z_\odot$, middle panel: 0.1 $Z_\odot$, right panel: 1.0 $Z_\odot$) with a different initial mass with a 23 $M_\odot$ BH companion in the orbital period from 0.04 - 40 days. Cross: initial RLOF, Square: case BA, Triangle: case BB, Diamond: case BC, circle: No MT.}
     \label{fig:interaction}
\end{figure*}

Figure \ref{fig:interaction} shows various interactions of a helium star with a 23 $M_\odot$ BH companion for different initial conditions (helium star's mass and metallicity, and the orbital period). Let us begin by describing the left panel of Figure \ref{fig:interaction} (i.e., 1.0 $Z_\odot$). Less massive helium stars have smaller radii at zamsHe but expand much faster after exhausting their core helium when compared with massive ones. First, when the initial orbital period is shorter than $\sim$ 0.1 days, massive helium stars ($\geqslant$ 6 $M_{\odot}$) are very likely to merge with their companions due to initial Roche-lobe overflow (RLOF), while less massive ones can have mass transfer during their early core helium-burning phase (Case BA). Second, it is shown that mass transfer occurs during the shell helium-burning phase (Case BB) for systems with $M_{\rm ZamsHe}$ $\leqslant$ 7.5 $M_{\odot}$ and $P_{\rm init.}$ $\leqslant$ 1.0 day. In addition, mass transfer during the core carbon-burning phase (Case BC) occurs for an initially higher orbital period and a slightly more massive helium star. For both Case BB and BC, the region of mass exchange extends with increasing initial mass and decreasing initial orbital period. At lower metallicities (middle and right panel), the outcomes are similar. Additionally, we note that the region for mass exchange slightly shrinks with decreasing metallicities. This is because stars at low metallicities tend to be more compact, which requires shorter initial orbital periods to interact with their companions.

\begin{figure*}
     \centering
     \includegraphics[width=1.0\textwidth]{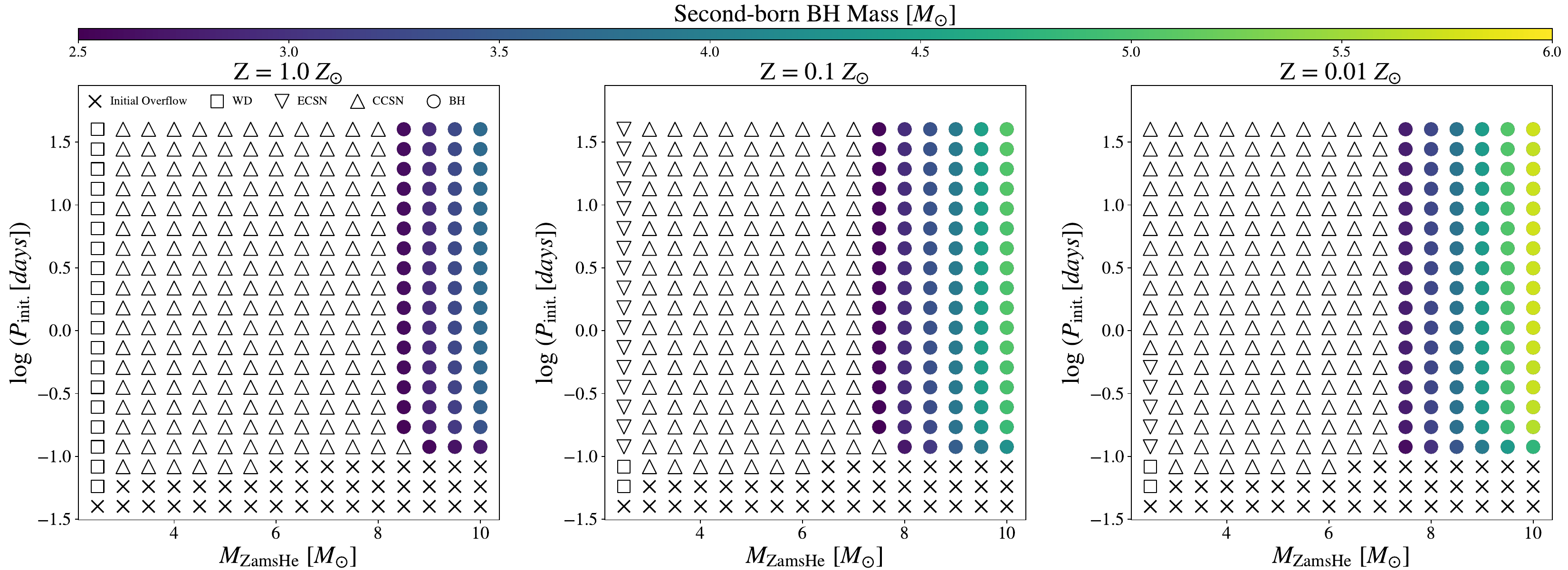}
     \caption{BH mass (the color bar) as a function of different initial helium star (Left panel: 1.0 $Z_\odot$, middle panel: 0.1 $Z_\odot$, right panel: 1.0 $Z_\odot$) masses and initial orbital periods. Square: white dwarf (WD), Triangle down: NS formed through ECSN, Triangle up NS formed through core collapse, Circle: BH.}
     \label{fig:BHmass}
\end{figure*}

After the carbon is depleted in the center, we assume that helium stars directly collapse to form compact objects by using the “delayed” supernova prescription \citep{Fryer2012}. Figure \ref{fig:BHmass} shows the BH mass as a function of initial helium star mass ($M_{\rm ZamsHe}$) and orbital period \footnote{In this work, we are exclusively focused on the BH binaries, leaving other binary compact objects for near future study.}. For helium stars reaching their carbon depletion in the center, we assume that electron-capture SN is formed with the CO core mass in the range between 1.37 $M_\odot$ \citep{Takahashi2013} and 1.43 $M_\odot$ \citep{Tauris2015}. In the left panel, we note that the lower limit mass of the helium star at solar metallicity that can form a BH is 8.5 $M_\odot$, below which a neutron star (NS) is formed instead (A helium star with $M_{\rm ZamsHe}$ $=$ 2.5 $M_\odot$ forms a white dwarf). For lower metallicities (0.1 $Z_\odot$ and 0.01 $Z_\odot$), the limit is shifted to 7.5 $M_\odot$. This is because stellar winds are strongly metallicity-dependent \cite[e.g.,][]{Vink2005}. 

\begin{figure*}
     \centering
     \includegraphics[width=1.0\textwidth]{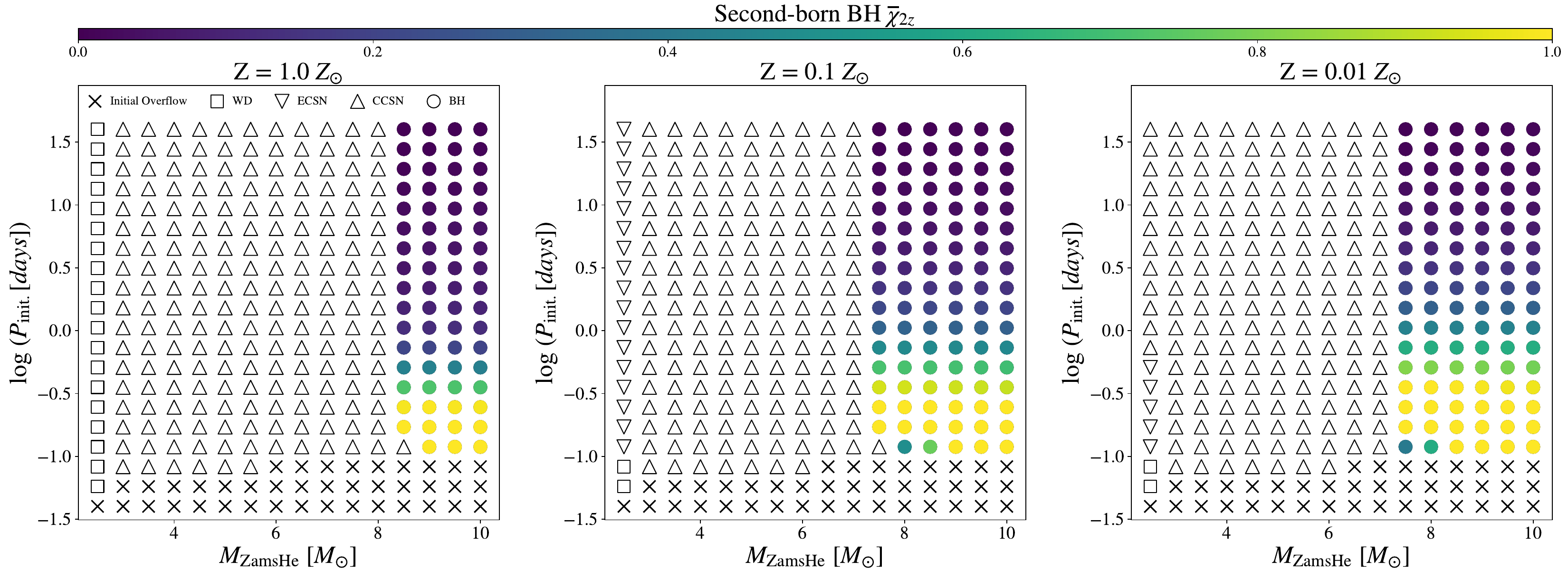}
     \caption{As in Figure \ref{fig:BHmass}, but the color bar refers to the BH spin, assuming the whole progenitor star directly collapses to form a BH without any mass or angular momentum loss. We also take into account the SN kicks imparted onto BHs, which could potentially change the angle between the spin direction at birth and the direction of the orbital angular momentum.}
     \label{fig:BHspin}
\end{figure*}

Similar to Figure \ref{fig:BHmass}, we present the BH spin ($\chi_{2z}$: the second-born BH's spin aligned to the direction of the orbital angular momentum) in various parameter spaces in Figure \ref{fig:BHspin}. Following the same method in \citet{Hu2022}, we take into account an SN kick that can potentially tilt the BH spin to the direction of the orbital angular momentum. The findings in the left panel show that high BH spin ($\chi_{2z}$ > 0.3) can be reached if an initial orbital period is shorter than $\sim$ 0.5 days ($\rm log$$P_{\rm init.} \sim$ - 0.3 days). At lower metallicities, stars tend to lose less angular momentum, which potentially results in higher BH spins with similar initial orbital periods. Interestingly, we note that two systems at initially lower metallicities, BH are formed with lower spins for short orbital periods ($P_{\rm init.} \sim$ 0.2 days). This is because the strong tides transfer the angular momentum from the helium star to the orbit, slowing down the progenitor and thus finally forming moderately rotating BHs.

\begin{figure*}
     \centering
     \includegraphics[width=1.0\textwidth]{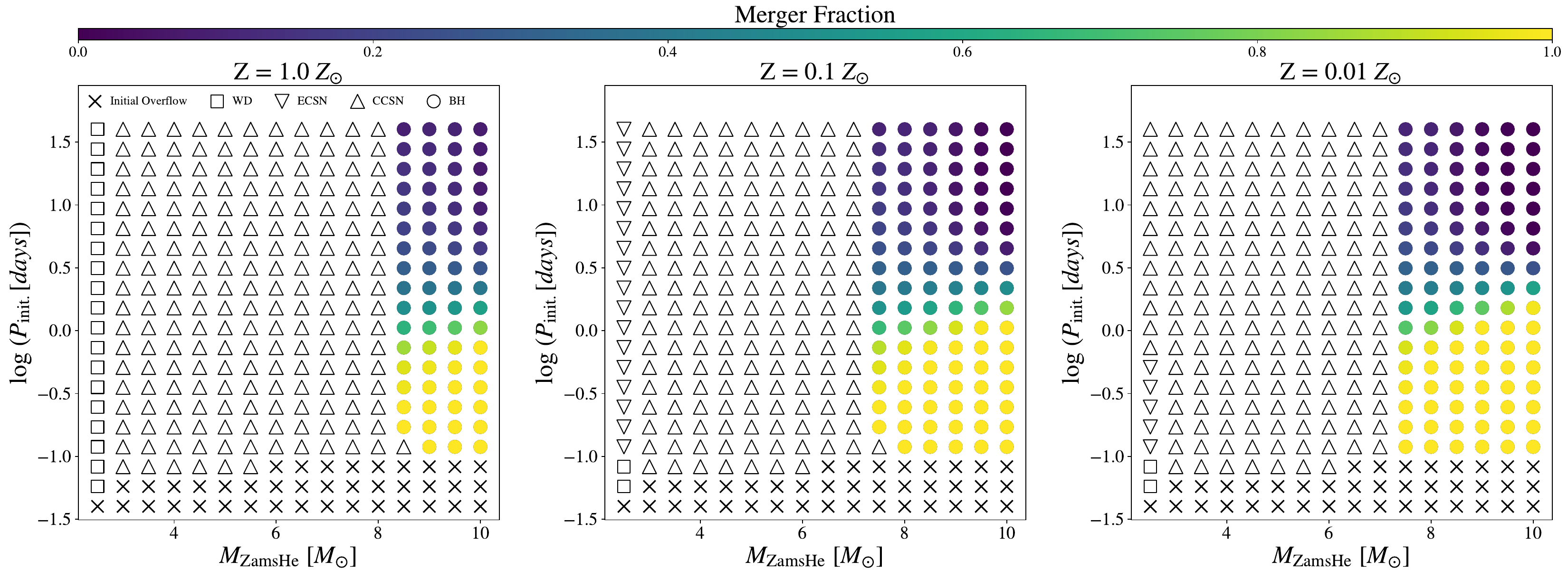}
     \caption{As in Fig.\ref{fig:BHmass}, but the color bar shows the merger fraction of binary systems.}
     \label{fig:mergerfraction}
\end{figure*}

\begin{figure*}
     \centering
     \includegraphics[width=1.0\textwidth]{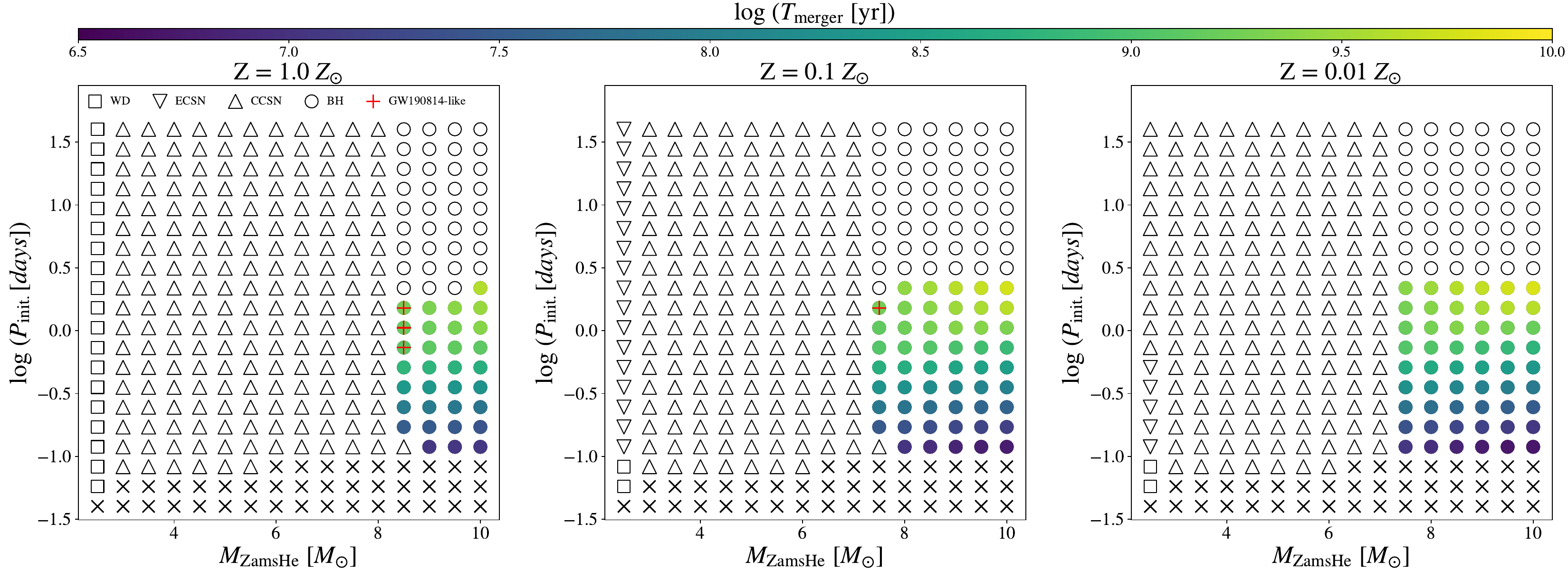}
     \caption{As in Fig.\ref{fig:BHmass}, but the color bar shows the merger time of BBH candidates due to GW emission. The red plus symbols represent systems that resemble GW190814-like. Note that the circles without filled color represent BBH systems with merger fractions less than 40\%.}
     \label{fig:tmerger}
\end{figure*}

The SN kicks imparted onto BHs are generally considered to modify the post-SN orbital separation and the direction of the BH's spin axis. Our calculations for natal kicks are based on the framework of \citet{Vicky1996,Wong2012} and more recent work in \citet{Callister2021}. We refer the readers to more detailed descriptions of our implementation in \cite{Hu2022}. We show in Figure \ref{fig:mergerfraction} the merger faction of BBHs for the parameter space considered in this work. Our findings show that the systems with a merger fraction larger than $\sim 50\%$ require the initial orbital period of $P_{\rm init.}$ shorter than $\sim$ 1.5 days ($\rm log$$P_{\rm init.} \sim 0.18$ days). For binaries surviving after the SN kick, we estimate the merger time of the BBH in Figure \ref{fig:tmerger} via gravitational-wave emission \citep{Peters1964} with an accurate fitting formula for eccentric binaries \citep{Mandel2021}. We here only consider binary systems surviving with the merger fraction $> 40\%$ after the SN kick. After combining the properties of mass and spin for the two BHs inferred earlier in this work, we thus obtain the candidates (red plus symbols in Figure \ref{fig:tmerger}) that probably resemble the event GW190814.

\section{Conclusions and Discussion}\label{sect5}
In this work, in the context of isolated binary evolution, we first identify that GW190814 is more likely to have been formed through the classical common envelope channel (CEE channel) when compared with the SMT or CHE channel. With the modified astrophysical priors on the effective $\chi_{\rm eff}$ predicted in \cite{Zevin2021}, we perform the Bayesian inference to obtain the updated properties of GW190814, which is consistent with the results in \cite{Abbott2020a}. With the newly-inferred properties of GW190814, we first perform a parameter space study with detailed binary modeling by taking into account various physical processes, including the stellar winds, tidal interaction, mass transfer, and SN kicks. Our findings show that the immediate progenitor of GW190814, just after the CE phase, is a BH of $\sim 23$ $M_{\odot }$ and a helium star of $M_{\rm ZamsHe}$ $\approx$ 8.5 $M_{\odot}$ at 1.0 $Z_{\odot}$ (7.5 $M_{\odot}$ at 0.1 $Z_{\odot}$) with an initial orbital period around 1.0 day. In addition, the metallicity for reproducing the formation history of GW190814-like events could not be lower than 0.1 $Z_{\odot}$. 

As shown in the earlier investigation \cite[see Channel B in Figure 6,][]{Zevin2020b}, GW190814-like events could be formed through the classical CE channel of two massive stars initially evolving at the metallicity of Z = 1/18 $Z_{\odot}$. It also shows that the immediate progenitor just after the CE phase is a binary system composed of a BH (3.0 $M_{\odot}$) and a helium star (23.4 $M_{\odot}$) in a very close orbit (0.02 AU). This short orbit corresponds to an orbital period of $P_{\rm orb}$ $\sim$ 0.2 days, where the progenitor of the BH will be efficiently spun up by tides \citep{Qin2018,Qin2019,Bavera2020,Hu2022,Hu2023,Qin2023}, resulting in a fast spinning BH. This result, however, seems not to be supported by the inferred spin magnitudes of the secondary component of GW190814. Consequently, we conclude that the metallicity of the progenitor for GW190814 is more likely to be higher than the one inferred in \citep{Zevin2020b}.

Our binary modeling could be impacted by some poorly understood physical processes, which mainly arise from stellar winds of helium stars and SN kicks imparted onto newly-formed BHs. Stronger winds tend to remove more mass and its corresponding angular momentum, which leads to forming less massive BHs with lower spins. Additionally, a larger SN kick could potentially cause a larger deviation of the newly-formed BH spin axis with respect to the direction of the orbital angular momentum, thus producing a smaller projected spin (e.g., $\Bar{\chi}_{2z}$ in Figure \ref{fig:BHspin}). Stellar wind mass loss is one of the paramount uncertain physical processes in the evolution of massive stars, and it can have a significant impact on the mass and the spin of resultant BHs. Recently, \cite{Tauris2022} argued that rather than a dynamical formation scenario, the isolated binary evolution can still explain the observed BBHs if BHs have their spin axis tossed by the SN kicks during the core collapse processes of massive stars. 

\section*{Acknowledgements}
We thank Christopher Berry for his helpful comments. The authors would like to thank the anonymous referee for her/his useful comments that improved the manuscript. Y.Q. acknowledges the support from the Doctoral research start-up funding of Anhui Normal University and from the Key Laboratory for Relativistic Astrophysics at Guangxi University. This work was partially supported by the National Natural Science Foundation of China (grant No. 12065017, 12192220, 12192221, U2038106) and the Natural Science Foundation of Universities in Anhui Province (grant No. KJ2021A0106). Q.W.T acknowledges support from the Natural Science Foundation of Jiangxi Province of China (grant No. 20224ACB211001). F.L. is supported by the Shanghai Post-doctoral Excellence Program and the National SKA Program of China (grant No. 2022SKA0130103). E.W.L is supported by the National Natural Science Foundation of China (grant No. 12133003). D.H.W is supported by the National Natural Science Foundation of China (NSFC) (grant No. 12103003).

\section*{Data availability}
The data generated in this work will be shared
upon reasonable request to the corresponding author.


\bibliographystyle{mnras}
\bibliography{ref} 


\label{lastpage}
\end{document}